# Surface dominated transport in single crystalline nanoflake devices of topological insulator $Bi_{1.5}Sb_{0.5}Te_{1.8}Se_{1.2}$

Bin Xia[+&], Peng Ren[+&], Azat Sulaev[+], Peng Liu[+], Shun-Qing Shen[#], Lan Wang[+*]

[+] *School of Physical and Mathematical Science, Nanyang Technological University, Singapore, 637371, Singapore*

[#] *Department of Physics, The University of Hong Kong, Pokfulam Road, Hong Kong, People's Republic of China*

&The two authors are equally contributed to this paper.

*Email: wanglan@ntu.edu.sg

We report experimental evidence of surface dominated transport in single crystalline nanoflake devices of topological insulator $Bi_{1.5}Sb_{0.5}Te_{1.8}Se_{1.2}$. The resistivity measurements show dramatic difference between the nanoflake devices and bulk single crystal. The resistivity and Hall analysis based on a two-channel model indicates that ~99% surface transport contribution can be realized in 200 nm thick BSTS nanoflake devices. Using standard bottom gate with $SiO_2$ as a dielectric layer, pronounced ambipolar electric field effect was observed in devices fabricated with flakes of 100 - 200 nm thick. Moreover, angle-dependent magneto-resistances of a nanoflake device with thickness of 596 nm are fitted to a universal curve for the perpendicular component of the applied magnetic field. The value of phase coherence length obtained from 2D weak antilocalization fitting further confirmed the surface dominated transport. Our results open a path for realization of novel electric and spintronic devices based on the topological helical surface states.

Topological insulators (TIs) are gapped bulk insulators with gapless Dirac surface states[1-4]. The surface states of these topological insulators are spin polarized and protected by time-reversal symmetry. A number of surface spectroscopy measurements, such as spin and angle-resolved photoemission spectroscopy (ARPES)[5-13] and scanning tunneling microscopy[14,15], have been used to detect the topologically nontrivial surface state in three-dimensional topological insulator $Bi_{1-x}Sb_x$, $Bi_2Se_3$, $Bi_2Te_3$, *etc*. The exotic surface states of topological insulators are expected to form a playground of various topological quantum effects and show great potential in spintronics and quantum computation[16-18]. To fulfill the expectations, realizing topological insulator systems with significant surface transport is essential. However, due to the defects or impurities in the samples, it is extremely difficult to eliminate the bulk contribution to electron transport[19-24]. Realizing surface dominated transport in current topological insulator systems is still a challenge despite extensive efforts involving chemical doping[22, 33-39], thin film or nanostructure fabricating[25-32] and electrical gating[21, 26, 32].

In this letter, we present strong evidences for surface dominated transport in nanoflake devices fabricated with several-hundred nanometer thick topological insulator $Bi_{1.5}Sb_{0.5}Te_{1.8}Se_{1.2}$ (BSTS). We performed electron transport measurements of both bulk single crystals and nanoflake devices of high quality single crystalline BSTS. Nanoflake devices shows a transition from semiconductor to metal near 100 - 150 K with decreasing temperature, while the bulk crystal shows semiconductor behavior in the measured temperature range from 300 K to 10 K and only present resistance saturation at very low temperature (< 40 K). At 10 K, the resistivity of the nanoflake devices decreases with the sample thickness and can get to hundred times smaller than that of the bulk single crystal. Bottom-gated devices fabricated with 150 - 200 nm thick BSTS nanoflakes show pronounced ambipolar electric field effect, which demonstrates significant topological surface transport. It was also found that the angle-dependent magneto-resistances of a 596 nm thickness nanoflake devices are fitted to a universal curve for the perpendicular component of an applied magnetic field. The phase coherence length obtained from a two-dimensional weak antilocalization fitting is much smaller than the sample thickness (596 nm), which clearly proves the 2D surface transport in the device. The temperature evolution of phase coherence length follows $T^{-0.47}$. All the experimental results suggest that surface dominated transport has been realized in several-hundred nanometer thick BSTS devices.

All the transport measurements were performed with the applied current in the (001) plane. The temperature dependence of the resistivity of the bulk single crystal and three nanoflake devices measured in zero field are shown in Fig. 1(a) and Fig. 1(b), respectively. The 102 μm thick bulk crystal shows semiconductor behavior in the measured temperature regime and reaches a saturation behavior when T < 40 K, which is due to the increasing transport contribution from surface[37, 38]. The resistivity at 10 K is more than 200 times larger than that at room temperature. Although the nanoflake device also shows the semiconductor behavior in the high temperature range, the transport characteristics transfer to a metallic type when T < 150 K. It should be noted from Fig. 1(a) and 1(b) that the resistivity of nanoflakes are hundred-times smaller than that of the bulk single crystal although the nanoflake is exfoliated from the bulk crystal used in the transport measurement (Fig. 1(a)). We have measured the sizes and R (T) curves of 10 nanoflake devices. As shown in the inset of Fig. 1(b), the resistivity of nanoflake devices at 10 K decreases with decreasing sample thickness. From the systematic decrease of resistivity with decreasing sample thickness and the energy-dispersive X-ray spectroscopy (EDS) as shown in supplementary information, we can conclude that the dramatic decrease of resistivity in nanoflake device is not due to chemical inhomogeneity. As the material system has already been proved to be a topological insulator[36, 37], the dramatic decrease of resistivity with decreasing sample thickness in nanoflake devices indicates that the contribution from the surface states plays a more and more important role as the thickness reduces. Suppose the total thickness of the two surface states is 5 nm, we use a simple model (supplementary information) to fit the resistivity of the surface states and bulk. The formula is shown below,

$$\rho = \frac{\rho_b \rho_s (t_b + t_s)}{t_s \rho_b + t_b \rho_s}, \qquad (1)$$

where $\rho$, $\rho_b$, $\rho_s$, $t_b$, $t_s$, are resistivity of the device, the resistivity of the bulk part, the resistivity of 5 nm thick surface states, the thickness of the bulk, and the thickness of the surface state (5nm), respectively. The fitting curve is shown as the red line in Fig. S2 and the inset of Fig. 1(b), which yields $\rho_b$ = 8.55 $\Omega\cdot$cm and $\rho_s$ = 2.27 × 10$^{-3}$ $\Omega\cdot$cm. For a device of 200 nm thick, the conductance contribution of the surface states is 98.9% at 10 K. As shown in the inset of Fig. 1(a), the Hall measurement of the single crystal is also fitted using a standard two-band model[32, 38] (details in supplementary information),

$$\rho_{xy}(B) = -\frac{B}{e} \frac{(n_b\mu_b^2 + n_s\mu_s^2/t) + B^2\mu_b^2\mu_s^2(n_b + n_s/t)}{(n_b\mu_b + n_s\mu_s/t)^2 + B^2\mu_b^2\mu_s^2(n_b + n_s/t)2} \quad (2)$$

where $\rho_{xy}$, $B$, $n_b$, $n_s$, $\mu_b$, $\mu_s$, and $t$ are the Hall resistivity, magnetic field, bulk charge density, surface charge density, bulk mobility, surface mobility, and sample thickness, respectively. The fitting yields $n_b = 1.1\times 10^{17}$ cm$^3$, $n_s = 2.5\times 10^{12}$ cm$^2$, $\mu_b = 25$ cm$^2$/V s, $\mu_s = 1767$ cm$^2$/V s. Based on the fitting results, the surface conductance contribution of a 200 nm thickness sample is 98.7%, which agrees well with the results obtained from resistance measurements.

If the electron transport is mainly due to the topological surface transport, we should be able to observe the ambipolar electric effect even in devices fabricated with hundred-nanometer thick BSTS nanoflake, which is indeed the case in our experiments. Fig. 2 shows the gate-voltage dependence of longitudinal resistance ($R_{xx}$) without applied magnetic field and Hall resistance ($R_{xy}$) at 1 Tesla magnetic field of a 172 nm thick device. To eliminate any possible non-symmetry effect of the sample and electrode contacts, all the Hall resistances were obtained from low field Hall measurement with magnetic field scanning from -1 Tesla to 1 Tesla. As the sample is 172 nm thick, the electrostatic gating can only shift the Fermi level of the bottom surface. The key finding is the resistance maxima appears near $V_g$ = -10 V. The resistance change near 35% with $\pm 60$V applied gate voltage. The quite sharp neutron point at 2 K and the gradual decrease of resistance with $V_g$ deviating from the neutron point indicate that the ambipolar behavior should be due to the Dirac cone of the topological surface state, which has been observed by ARPES[36]. If it was due to the normal band banding, there would be a wide constant resistance region when the Fermi level in the band gap and much dramatic resistance decrease when the Fermi level enters into conduction or valence band. The Hall resistance also changes with applied gate voltage as shown in Fig. 2(a), which further confirms the ambipolar electric gating effect in this device. From the measured Hall resistance $R_H$ (150 Ω/T at 2 K), if we neglect the Hall contribution from the bulk, the charge density can be estimated to be about $4.2 \times 10^{12}$ cm$^{-2}$ (bottom + top), which approximately agrees with the bulk Hall fitting results as aforementioned. For a 300 nm SiO$_2$ dielectric layer, a gate voltage of 60 volts can induces n = $4.4 \times 10^{12}$ cm$^{-2}$, therefore the 300 nm SiO$_2$ dielectric layer should be able to generate ~50% resistance change because we can only tune the bottom surface. However, considering the 176 nm thick side surface of the devices, which can contribute ~5 - 10% surface conductance contribution, and the small amount of charge from the bulk, it is very reasonable that we only get 30 - 40 % resistance change.  It should be pointed out that we have observed similar ambipolar behavior in many devices fabricated with several hundred nanometer thick BSTS. The results of several devices are shown in the supplementary information. Fig. 2(b) shows the gate voltage dependence of resistance at various temperatures. It is observed that the shape of the $R(V_g)$ curves become more broad with increasing temperature, which is a standard ambipolar behavior as shown in graphene[40]. It also agrees well with the deceasing surface conductance contribution with increasing temperature. The shift of resistance peaks with increasing temperature indicates the shift of chemical potential of the bottom surface with increasing temperature.

To further confirm the surface dominated electron transport in the nanoflake device, we have performed the angle-dependent magnetoresistance measurements using a very thick nanoflake device (596 nm, Fig. S6 in supplementary information). To eliminate any effect from Hall resistance, all the magnetoresistance measurements were performed from -2 Tesla to 2 Tesla and carried out corresponding calculation process to obtain intrinsic magnetoresistance. The magnetoresistance was measured by tilting the bulk crystal and nanoflake device with respect to the applied magnetic field from 0° to 90°. $\theta = 0°$ means the magnetic field parallel to the (001) surface while the $\theta = 90°$ means the magnetic field perpendicular to the (001) surface. Fig. 3(a) and 3(b) show the field dependence of magnetoconductances $\Delta G = G(B) - G(0)$ with various θ for the bulk single crystal and nanoflake device at 2 K, respectively. Both the bulk crystal and nanodevice show weak antilocalization behavior in the low magnetic field region. Such a weak antilocalization at low temperature has been attributed to both strong spin-orbital coupling and topological π Berry phase of two dimensional surface states[41]. It is very obvious that ΔG decreases with decreasing θ for both bulk single crystal and nanoflake device and it is more pronounced for the nanoflake device. The variation of magnetoconductance as a function of the perpendicular component of the magnetic field for the bulk single crystal and nanoflake device at 2 K, are shown in Fig. 3(c) and 3(d), respectively. The *G vs.* $B\sin\theta$ (the perpendicular component of the applied field) curves of the nanoflake device are perfectly merged into one universal curve. It should be noticed that a misalignment between the sample and applied magnetic field has a large effect when the θ is near 0 while it has a negligible effect when θ is a large value, which can be inferred from the small different between the θ = 80° and θ = 90° curves and the large difference between θ = 0° and θ = 5° curves. Based on this, we speculate that the very small magnetoconductance at θ = 0 may be due to the small unavoidable misalignment between the sample and the applied magnetic field in experiment. Therefore, the perfect fitting into a universal curve for all θ value (except θ = 0) as shown in Fig. 3(d) is a support of the conclusion obtained from the resistivity and Hall measurements, surface dominated transport (96.9% in a 596 nm device). For the bulk single crystal, it is not a surprise that the weak antilocalization is observed for all the angles and $G$ vs. $B\sin\theta$ curve cannot be fitted to one universal curve because of the coexistence of both the bulk spin-orbit coupling effects and helical surface state contribution.

Magnetoconductances $\Delta G$ at different temperatures (2 K, 4 K, 7 K, 10 K, 15 K, 35 K, and 45 K) have been obtained in a magnetic field perpendicular to the (001) plane (θ = 90°), as shown in Fig. 4(a) and 4(b) for bulk single crystal and nanoflake device, respectively. The Hikami-Larkin-Nagaoka (HLN) formula

$$\Delta\sigma_{2D} = \left[-\frac{\alpha e^2}{2\pi^2\hbar}\right]\left[\ln(\frac{\hbar}{4eL_\varphi^2 B}) - \Psi(\frac{1}{2} + \frac{\hbar}{4eL_\varphi^2 B})\right]$$

is used to fit the magnetoconductance observed in both bulk and nanoflake device, where $\Delta\sigma_{2D}$ is the two dimensional conductivity ($\Delta\sigma_{2D} = \frac{\Delta G \cdot L}{W}$, L and W are the length and width of transport channel, respectively), $\Psi$ is the digamma function, $L_\varphi$ is the phase coherence length, and α is a prefactor which contains information about the nature of the electrons in topological insulators[42]. The fitted curves are plotted in Fig. 4(a) and 4(b) in solid lines. Due to the three dimensional bulk contribution, the fitting curves do not agree well with experimental data for bulk single crystal samples. For the nanoflake device, the two dimensional fitting curves agree very well with the experimental data for both fitting range from 0 to 2 Tesla and from 0 to 0.5 Tesla at all temperatures from 2 K to 45 K. The fitting results of $\alpha$ and $L_\varphi$ of nanoflake device in both fitting ranges are shown in Fig. 4(c) and 4(d) respectively. It is clear that similar results are obtained from different fitting ranges. The power law fit gives $L_\varphi \propto T^{-0.47}$ (2 Tesla) and $L_\varphi \propto T^{-0.44}$ (0.5 Tesla), which indicates the two-dimensional transport characteristics of the device[43]. It should be noted that the fitted $L_\varphi$ at all temperatures are much smaller than the thickness of the sample (596 nm), which strongly indicates that the two dimensional electron transport characteristics are due to the surface states. From this point of view, our experiment is very different from previous measurements on ultra-thin $Bi_2Se_3$ and $Bi_2Te_3$ films. The two dimensional transport behavior in ultra-thin $Bi_2Se_3$ and $Bi_2Te_3$ films cannot be completely attributed to the surface states, because the $L_\varphi$ is larger than the thickness of the film and therefore the film itself, including the bulk and surface states, forms a two dimensional transport system. Since the thickness (596 nm) of our nanoflake BSTS device is much larger than $L_\varphi$ ( ~180 nm at 2 K, ~110 nm at 10 K, and ~60 nm at 30 K), it is a three dimensional transport system. The two dimensional transport behavior of weak antilocalization can only originate from the helical surface states. The coefficient $\alpha$ takes a value of -1/2 for a traditional 2D electron system with strong spin-orbit coupling and one helical surface with a single Dirac cone. Since carriers on both up and bottom surface can contribute the conduction in topological insulator samples, the ideal value of $\alpha$ is -1. As shown in Fig. 4(c), the perfect fitting using HLN formula generates $\alpha$ values between -0.7 to -0.8 at various temperatures, which also supports the 2D surface transport.

In conclusion, we demonstrate surface dominated transport in single crystalline nanoflake devices of topological insulator BSTS. The resistivity and Hall analysis based on a two-channel model indicates that ~99% of the surface transport contribution can be realized in 200 nm thick BSTS devices. The pronounced electric gated ambipolar behavior proves the topological surface transport. Moreover, the angle and temperature dependent weak antilocalization effect of bulk single crystals and nanoflake devices strongly suggest the surface dominated transport in nanoflake device, which is further confirmed by the fact that the phase coherence length $L_\varphi$ obtained by 2D HNL fitting is much smaller than the thickness of device.

**Supporting Information Available**

The two-channel model, synthesis and characterization of BSTS single crystal, nanoflake device fabrication, time dependent universal conductance fluctuation, and more transport results. The material is available free of charge via the Internet at http://pubs.acs.org

**AKNOWLEDGMENTS**

Support for this work came from Singapore National Research Foundation (RCA-08/018), MOE Tier 2 ( MOE2010-T2-2-059) and the Research Grant Council of Hong Kong under Grant No. HKU705110P.

Figure captions:

**Figure 1.** Temperature dependence of resistivity in zero field of (a) bulk single crystal with a thickness of 102 μm and (b) nanoflake devices with various thicknesses. The inset of (a) shows the Hall measurement results of the 102 μm thick BSTS single crystal and the fitting (the red curve) based on a two-channel model. The inset of (b) shows the resistivity values of nanoflake devices with various thicknesses. The red line is the fitting curve based on the two-channel model.

**Figure 2.** (a) The gate voltage dependence of resistance ($R_{xx}$) and Hall resistance ($R_{xy}$) at 1 Tesla filed of a BSTS nanoflake device at 2 K. (b) The temperature evolution of the $R_{xx}(V_g)$ curve.

**Figure 3.** Angle-dependent conductance of a bulk BSTS single crystal and a nanodevice fabricated with a 596 nm thick BSTS flake. θ is the angle between the direction of the magnetic field and the current flow plane. (a) and (b) show the magnetoconductance of bulk BSTS single crystal and nanoflake device at 2 K, respectively. (c) and (d) show the curves of magnetoconductance *vs.* the perpendicular component of the magnetic field of bulk BSTS single crystal and nanoflake device at 2 K, respectively.

**Figure 4.** Temperature-dependent magnetoconductance of the BSTS single crystal and 596 nm thick nanoflake device with θ = 90°. (a) Magnetoconductance of the bulk single

crystal at different temperatures. The HNL fitting curves are plotted in solid lines. (b) Magnetoconductance of the nanoflake device at different temperatures. The HNL fitting curves are plotted in solid lines. (c) The fitted values of α at different temperatures for the nanoflake device in the fitting range of 0 - 2 Tesla and 0 - 0.5 Tesla, respectively. (d) The fitted values of $L_\varphi$ at different temperatures of the nanoflake device in the fitting range of 0 - 2 Tesla and 0 - 0.5 Tesla, respectively. The fitting values follow the power law of temperature $T^{-0.47}$ and $T^{-0.43}$, respectively

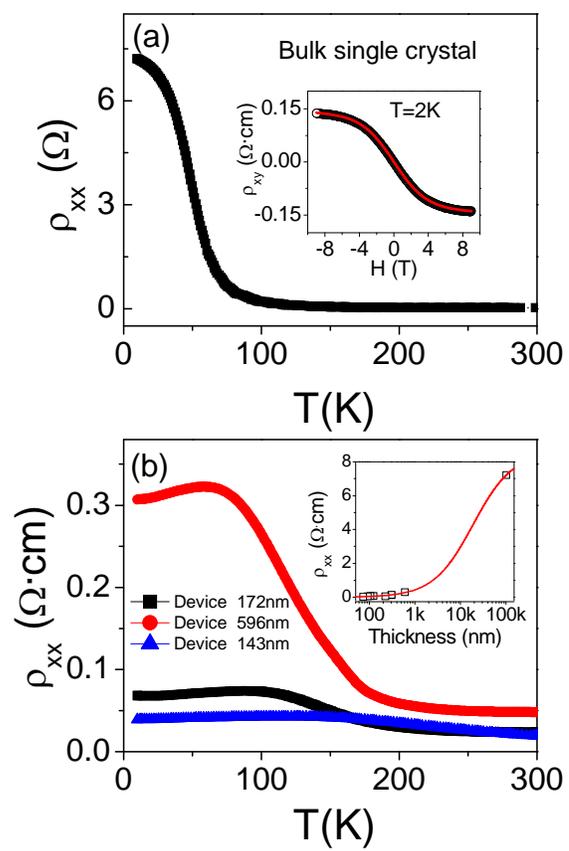

Figure 1

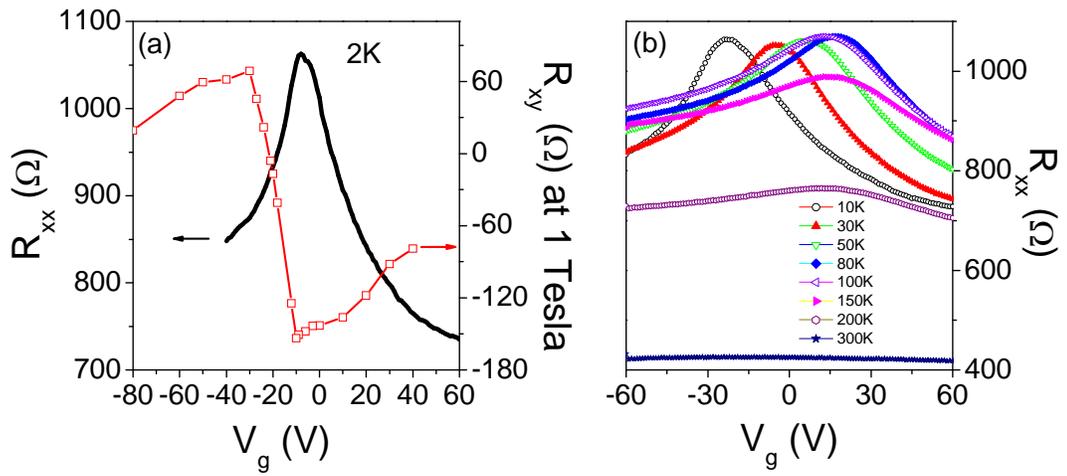

Figure 2

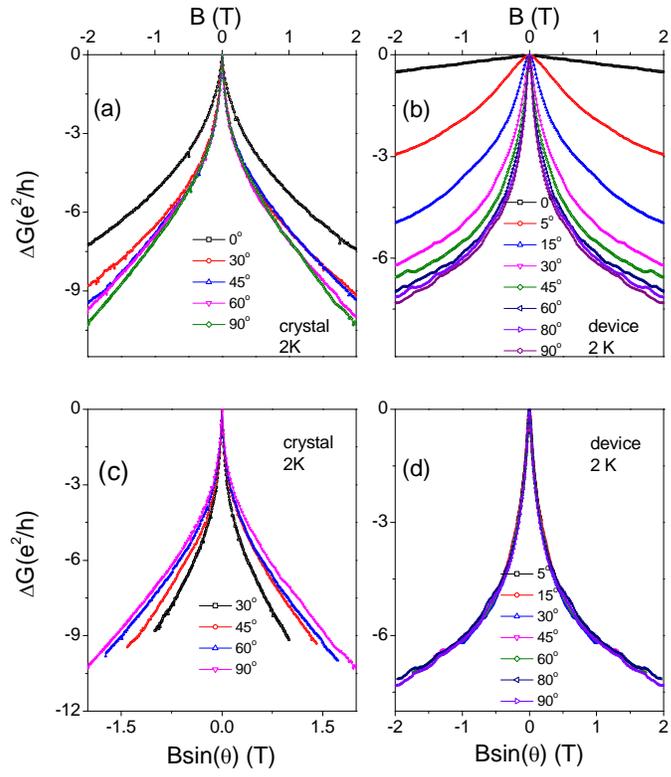

Figure 3

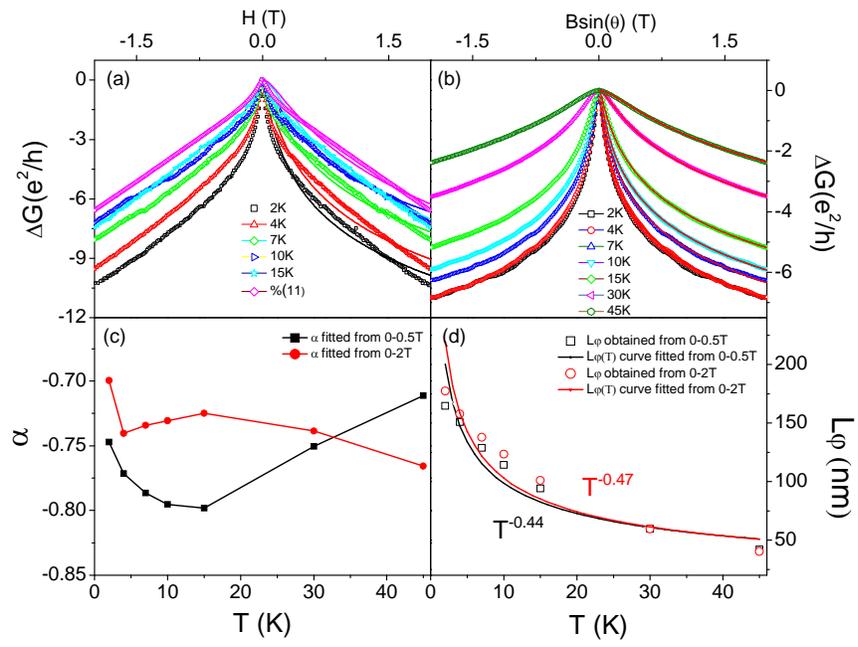

Figure 4

# REFERENCES

(1) Fu, L; Kane, C. L.; Mele, E. J. *Phys. Rev. Lett.* **2007**, 98, 106803.

(2) Moore J. E.; *Nature* **2010**, 464, 194.

(3) Qi, X. L.; Zhang, S. C. *Phys. Today* **2010,** 63, 33

(4) Hasan, M.Z.; Kane, C. L. *Rev. Mod. Phys.* **2010**, 82, 3045.

(5) Xia, Y.; Dian, D.; Hsieh, D.; Wray, L.; Pal, A.; Lin, H.; Bansil, A.; Grauer, D.; Hor, Y. S.; Cava, R. J.; Hasan, M. Z. N*at. Phys.* **2009**, 5, 398.

(6) Chen, Y. L.; Analytis, J. G.; Chu, J. H.; Liu, Z. K.; Mo, S. K.; Qi, X. L.; Zhang, H. J.; Lu, D. H.; Dai, X.; Fang, Z.; Zhang, S. C.; Fisher, I. R.; Hussain, Z.; and Shen, Z. X. *Science* 2**009**, 325, 178.

(7) Hsieh, D.; Xia, Y.; Wray, L.; Qian, D.; Pal, A.; Dil, J. H.; Osterwalder, J.; Meier, F.; Bihlmayer, G.; Kane C. L.; Hor, Y. S.; Cava, R. J.; Hasan, M. Z. *Science* **2009**, 323, 919.

(8) Hirahara, T.; Sakamoto, Y.; Takeichi, Y.; Miyazaki, H.; Kimura, S. I.; Matsuda, I.; Kakizaki, A.; Hasegawa, S. *Phys. Rev. B* **2010**, 82, 155309.

(9) Jozwiak, C; Chen, Y. L.; Fedorov, A. V.; Analytis, J. G.; Rotundu C. R.; Schmid, A. K.; Denlinger, J. D.; Chuang, Y. D.; Lee, D. H.; Fisher, I. R.; Birgeneau, R. J.; Shen, Z. X.; Hussain, Z.; Lanzara, A.; *Phys. Rev. B* **2011**, 84, 165113.

(10) Hsieh, D.; Xia, Y.; Qian, D.; Wray, L.; Dil, J. H.; Meier, F.; Osterwalder, J.; Patthey, L.; Checkelsky, J. G.; Ong, N. P.; Fedorov, A. V.; Lin, H.; Bansil, A.; Grauer, D.; Hor, Y. S.; Cava, R. J.; Hasan, M. Z. *Nature* **2009**, 460, 1101.

(11) Nishide, A.; Taskin, A. A.; Takeichi, Y.; Okuda, T.; Kakizaki, A.; Hirahara, T.; Nakatsuji, K.; Komori, F.; Ando, Y.; Matsuda, I. *Phys. Rev. B* **2010**, 81, 041309.

(12) Souma, S.; Kosaka, K.; Sato, T.; Komatsu, M.; Takayama, A.; Takahashi, T.; Kriener, M.; Segawa, K.; Ando, Y. *Phys. Rev. Lett.* **2011**, 106, 216803.

(13) Xu, S. Y.; Wray, L. A.; Xia, Y.; von Rohr, F.; Hor, Y. S.; Dil, J. H.; Meier, F.; Slomski, B.; Osterwalder, J.; Neupane, M.; Lin, H.; Bansil, A.; Fedorov, A.; Cava, R. J.; Hasan, M. Z. **2011**, arXiv:1101.3985.

(14) Alpichshev, Z.; Analytis, J. G.; Chu, J. H.; Fisher, I. R.; Chen, Y. L.; Shen, Z. X.; Fang, A.; A. Kapitulnik, Phys. Rev. Lett. **2010**, 104, 016401.

(15) Zhang, T.; Cheng, P.; Chen, X.; Jia, J.-F.; Ma, X. C.; He, K.; Wang, L. L.; Zhang, H. J.; Dai, X.; Fang, Z.; Xie, X. C.; Xue, Q. K. *Phys. Rev. Lett.* **2009**, 103, 266803.

(16) Fu L.; Kane, C. L. *Phys. Rev. Lett.* **2008**, 100, 096407.

(17) Qi, X. L.; Li, R.; Zang, J.; Zhang, S. C.; *Science,* **2009**, 323, 1184.

(18) Yu, R.; Zhang, W.; Zhang, H. J.; Zhang, S. C.; Dai, X.; Fang, Z. *Science* **2010**, 329, 61.

(19) Analytis, J. G.; Chu, J. H.; Chen, Y.; Corredor, F.; McDonald, R. D.; Shen, Z. X.; Fisher, I. R. *Phys.Rev. B* **2010,** 81, 205407.

(20) Hor, Y. S.; Richardella, A.; Roushan, P.; Xia, Y.; Checkelsky, J. G.; Yazdani, A.; Hasan, M. Z.; Ong, N. P.; Cava, R. J.; *Phys. Rev. B* **2009**, 79, 195208.

(21) Checkelsky, J. G.; Hor, Y. S.; Cava, R. J.; Ong, N. P. *Phys. Rev. Lett.* **2010**, 106, 196801.

(22) Checkelsky, J. G.; Hor, Y. S.; Liu, M.-H.; Qu, D.-X.; Cava, R. J.; Ong, N. P. *Phys. Rev. Lett.* **2009**, 103, 246601.

(23) Butch, N. P.; Kirshenbaum, K.; Syers, P.; Sushkov, A. B.; Jenkins, G. S.; Drew, H. D.; Paglione, J. *Phys. Rev. B* **2010**, 81, 241301.

(24) Qu, D. X.; Hor, Y. S.; Xiong, J.; Cava, R. J.; Ong, N. P.; Science, **2010,** 329, 821

(25) Peng, H.; Lai, K.; Kong, D.; Meister, S.; Chen, Y.; Qi, X.-L.; Zhang, S.-C.; Shen, Z.-X.; Cui, Y.

Nat. Mater. **2010**, 9, 225.

(26) Chen, J.; Qin, H. J.; Yang, F.; Liu, J.; Guan, T.; Qu, F. M.; Zhang, G. H.; Shi, J. R.; Xie, X. C.; Yang, C. L.; Wu, K. H.; Li, Y. Q.; Lu, L. *Phys. Rev. Lett.* **2010**, 105, 176602.

(27) Kim, Y. S.; Brahlek, M.; Bansal, N.; Edrey, E.; Kapilevich, G. A.; Iida, K.; Tanimura, M.; Horibe, Y.; Cheong, S. W.; Oh, S. *Phys. Rev. B* **2011**, 84, 073109.

(28) Liu, M.; Zhang, J.; Chang, C. Z.; Zhang, Z.; Feng, X.; Li, K.; He, K.; Wang, L. L.; Chen, X.; Dai, X.; Fang, Z.; Xue, Q. K.; Ma, X.; Wang, Y.; *Phys. Rev. Lett.* **2012**, 108, 036805.

(29) Liu, M.; Chang, C. Z.; Zhang, Z.; Zhang, Y.; Ruan, W.; He, K.; Wang, L. L.; Chen, X.; Jia, J. F.; Zhang, S. C.; Xue, Q. K.; Ma, X.; Wang, Y. *Phys. Rev. B* **2011**, 83, 165440.

(30) He, H. T.; Wang, G.; Zhang, T.; Sou, I. K.; Wong, G. K. L.; Wang, J. N.; Lu, H. Z.; Shen, S. Q.; F. C. Zhang, *Phys. Rev. Lett.* **2011**, 106, 166805.

(31) Matsuo, S.; Koyama, T.; Shimamura, K.; Arakawa, T.; Nishihara, Y.; Chiba, D.; Kobayashi, K.; Ono, T.; Chang, C. Z.; He, K.; Ma, X. C.; Xue, Q. K. *Phys. Rev. B* **2012**, 85, 075440.

(32) Steinberg, H.; Gardner, D. R.; Lee Y. S.; Jarillo-Herrero, P. *Nano Lett.* **2010**, 10, 5032.

(33) Ren, Z.; Taskin, A. A.; Sasaki, S.; Segawa, K.; Ando, Y.; *Phys. Rev. B* **2010**, 82, 241306.

(34) Jia, S.; Ji, H.; Climent-Pascual, E.; Fuccillo, M. K.; Charles, M. E.; Xiong, J.; Ong, N. P.; Cava, R. J. *Phys. Rev. B* **2011**, 84, 235206.

(35) Xiong, J.; Petersen, A. C.; Qu, D.; Cava, R. J.; Ong, N. P. **2011** arXiv:1101.1315.

(36) Arakane, T.; Sato, T.; Souma, S.; Kosaka, K.; Nakayama, K.; Komatsu, M.; Takahashi, T.; Ren, Z. ; Segawa, K.; Ando, Y. Nat. Comm. **2012**, 3, 636.

(37) Taskin, A. A.; Ren, Z.; Sasaki, S.; Segawa, K.; Ando, Y. Phys. Rev. Lett. **2011**, 107, 016801.

(38) Ren, Z.; Taskin, A. A.; Sasaki, S.; Segawa, K.; Ando, Y. Phys. Rev. B **2011**, 84, 075316.

(39) Hong, S. S.; Chua, J. J.; Kong, D. S.; Cui, Y. Nat. Comm. **2012**, **3**, 757.

(40) Novoselov, K. S.; Geim, A. K.; Morozov, S. V.; Jiang, D.; Zhang, Y.; Dubonos, S. V.; Grigorieva, I. V.; Firsov, A. A. *Science* **2004**, 306, 666.

(41) Lu, H. Z.; Shen, S. Q. *Phys. Rev. B* **2011,** 84, 125138.

(42) McCann, E.; Kechedzhi, K.; Fal'ko, V. I.; Suzuura, H.; Ando, T.; Altshuler, B. L. *Phys. Rev. Lett.* **2006**, 97**,** 146805.

(43) Santhanam, P.; Wind, S.; Prober, D. E. *Phys. Rev. Lett.* **1984**, 53, 1179.

# Supplementary information

## I. Two-channel model for calculating the sample resistivity

To calculate the thickness dependence of the resistivity of BSTS single crystal, a simple two-channel model is utilized.

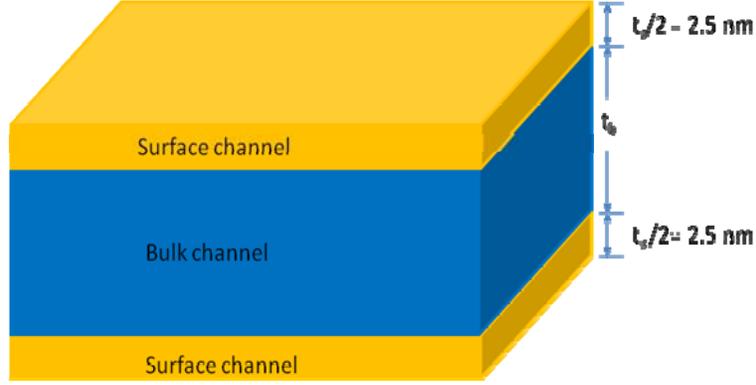

Fig. S1. Schematic diagram of a two-channel model for resistivity of BSTS nanoflake device.

$R_b$, $R_s$, $\rho_b$, $\rho_s$, $l$, $w$, $t_b$, $t_s = 0.5$ nm are the bulk resistance, surface resistance, bulk resistivity, surface resistivity, sample length, sample width, the thickness of the bulk channel, the thickness of the surface channel, respectively.

$$R_b = \frac{\rho_b \cdot l}{w \cdot t_b}, \tag{S1}$$

$$R_s = \frac{\rho_s \cdot l}{w \cdot t_s}, \tag{S2}$$

$$R = \frac{1}{\frac{1}{R_b} + \frac{1}{R_s}} = \frac{1}{\frac{w \cdot t_b}{\rho_b \cdot l} + \frac{w \cdot t_s}{\rho_s \cdot l}} = \frac{\rho_s \cdot \rho_b \cdot l}{(t_s \cdot \rho_b + t_b \cdot \rho_s) \cdot w}, \tag{S3}$$

$$\rho = \frac{R \cdot w \cdot (t_b + t_s)}{l} = \frac{\rho_b \rho_s (t_b + t_s)}{t_b \rho_s + t_s \rho_b}. \tag{S4}$$

To accurately determine the thickness and width of the nanoflake devices, atomic force microscopy (AFM) and scanning electron microscopy (SEM) were used, respectively. The microscopy was performed after the transport measurements.

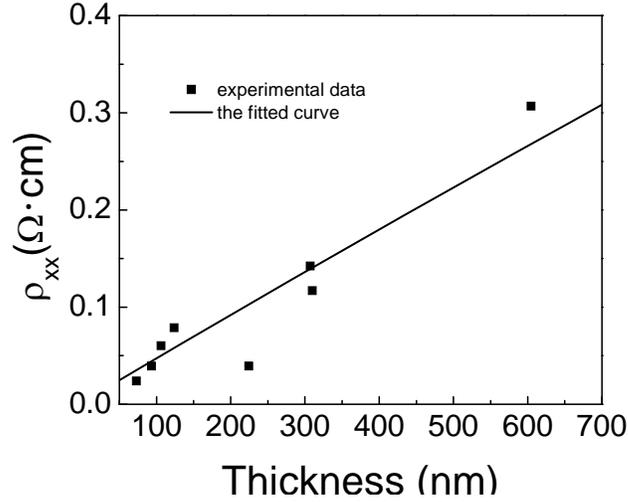

Fig. S2. The zoom-in of the inset of Fig. 1(b).

As stated in the paper, the fitting yields $\rho_b$ = 8.55 $\Omega\cdot$cm and $\rho_s$ = 2.27 × $10^{-3}$ $\Omega\cdot$cm. The linear scale zoom-in figure for nanodevices (thickness < 600 nm) is shown Fig. S2.

The conductance contribution of the surface state can be easily calculated as

$$\sigma_b = \frac{W \cdot t_b}{\rho_b \cdot l} \qquad \sigma_s = \frac{W \cdot t_s}{\rho_s \cdot l} \tag{S5}$$

$$\frac{\sigma_s}{\sigma_s + \sigma_b} = \frac{\frac{W \cdot t_s}{\rho_s \cdot l}}{\frac{W \cdot t_s}{\rho_s \cdot l} + \frac{W \cdot t_b}{\rho_b \cdot l}} = \frac{1}{1 + \frac{t_b \cdot \rho_s}{\rho_b \cdot t_s}} \tag{S6}$$

Suppose $t_b$ = 200 nm, $t_s$ = 5 nm, the surface contribution can be easily calculated. The value is 98.9% if $\rho_b$ = 8.55 $\Omega\cdot$cm and $\rho_s$ = 2.27 × $10^{-3}$ $\Omega\cdot$cm.

## II. Two-channel model for calculating Hall effect,

We use Drude model
$$m\frac{d\vec{v}}{dt} = -e(\vec{E} + [\vec{v} \times \vec{B}]) - \frac{m\vec{v}}{\tau},$$ (S7)

$$\vec{j} = ne\vec{v}.$$ (S8)

At steady state
$$e\vec{E} = -e[\vec{v} \times \vec{B}]) - \frac{m\vec{v}}{\tau},$$ (S9)

$$\vec{E} = -\frac{1}{ne}[ne\vec{v} \times \vec{B}]) - ne\vec{v} \cdot \frac{m}{ne^2\tau}.$$ (S10)

Define
$$R_H = \frac{1}{ne}, \quad \text{and} \quad \rho = \frac{m}{ne^2\tau},$$ (S11)

We obtain
$$\vec{E} = \rho \cdot \vec{j} - R_H(\vec{j} \times \vec{B}).$$ (S12)

When the magnetic field $\vec{B}$ points to z direction, we obtain

$$\begin{pmatrix} E_x \\ E_y \end{pmatrix} = \begin{pmatrix} \rho & -R_H H \\ R_H H & \rho \end{pmatrix} \begin{pmatrix} j_x \\ j_y \end{pmatrix}.$$ (S13)

Then the conductivity
$$\hat{\sigma} = \begin{pmatrix} \rho & -R_H H \\ R_H H & \rho \end{pmatrix}^{-1} = \frac{1}{\rho^2 + (R_H H)^2} \begin{pmatrix} \rho & R_H H \\ -R_H H & \rho \end{pmatrix}.$$ (S14)

For more than one channel transport,
$$\vec{j} = \sum_i \hat{\sigma}_i \vec{E}.$$ (S15)

Using
$$\frac{1}{\rho} = ne\mu \quad \text{and} \quad \frac{R_H}{\rho} = \mu,$$ (S16)

We obtain
$$\hat{\sigma} = \begin{pmatrix} \sum_i \dfrac{ne\mu_i}{1+\mu_i^2 B_i^2} & \sum_i \dfrac{ne\mu_i^2}{1+\mu_i^2 B_i^2} \\ \sum_i \dfrac{ne\mu_i^2 B}{1+\mu_i^2 B^2} & \sum_i \dfrac{ne\mu_i^2}{1+\mu_i^2 B^2} \end{pmatrix}. \quad (S17)$$

Using
$$\hat{\rho} = \hat{\sigma}^{-1}, \quad (S18)$$

For a two-channel system, we obtain

$$\rho_{xy}(B) = -\frac{B}{e}\frac{(n_1\mu_1^2 + n_2\mu_2^2) + B^2\mu_1^2\mu_2^2(n_1+n_2)}{(n_1\mu_1 + n_2\mu_2)^2 + B^2\mu_1^2\mu_2^2(n_1+n_2)^2} \quad (S19)$$

The formula is for two three-dimensional transport channel. If channel 2 is a two dimensional transport channel (thickness = 0), the formula should be written as

$$\rho_{xy}(B) = -\frac{B}{e}\frac{(n_b\mu_b^2 + n_s\mu_s^2/t) + B^2\mu_b^2\mu_s^2(n_b + n_s/t)}{(n_b\mu_b + n_s\mu_s/t)^2 + B^2\mu_b^2\mu_s^2(n_b + n_s/t)^2}, \quad (S20)$$

which is the Eq. (2) in the main text. Here, $t$ is the sample thickness.

Eq. S18 is clear for understanding physics concept. However, $n_b$, $n_s$, $\mu_b$, and $\mu_s$ are very large numbers, it is difficult to use Eq. S18 for curve fitting. By using Eq. S14 or derive from Eq. S12 directly, we transfer Eq. S18 to another equation,

$$\rho_{xy}(B) = -\frac{(R_b\rho_s^2 + R_s\rho_b^2) + B^2 R_b^2 R_s^2 (R_b + R_s)}{(\rho_b + \rho_s)^2 + B^2(R_b + R_s)^2}. \quad (S21)$$

Here, $R_b$ and $\rho_b$ are the Hall coefficient and resistivity of the bulk electrons. $R_s = t/en_s$ and $\rho_s = \rho_{sheet}/t$, where $\rho_{sheet}$ is the surface sheet resistance.

## III. Single crystal growth and characterization

High quality BSTS single crystals were grown using modified Bridgeman methods. High purity (99.9999%) Bi, Sb, Te, and Se with a molar ratio of 1.5:0.5:1.8:1.2 were first thoroughly mixed and then reacted at 950°C for 1 week in an evacuated quartz tube in a box furnace. We then located the quartz tube vertically in a specially designed furnace with large temperature gradient. The temperature is then decreased to room temperature over 3 weeks, with different cooling speed in different temperature regions. The obtained crystals are easily cleaved and revealed a flat and big shiny surface as shown in the inset of Fig. S4(a).

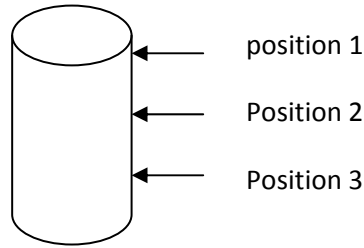

Fig. S3. A schematic diagram of an as grown single crystal.

As shown in Fig. S3, an as grown single crystal is a cylinder with ~0.7 cm$^2$ (the cross section area) × 2 cm (height). Large cleaved crystals can be cut from different positions of the crystal.

The X-ray diffraction pattern indicates the high quality of our samples. Fig. S4(a) shows the wide angle x-ray diffraction from a bulk crystal oriented with the scattering vector perpendicular to the (001) family of planes. No peaks from other plane families can be observed. It should be noted that the count of the (006) peak is larger than 300 k. Only high quality single crystals can get to this value in XRD facility. The inset of Fig. S4(a) shows the large and shining (001) surface of the single crystal.

Energy dispersive X-ray spectroscopy (EDS) was employed to probe the homogeneity of the samples. The small area EDS mapping as shown in Fig. S4(b) indicates the element distribution in a small area is homogeneous. We also performed the EDS measurements at various points of the bulk single crystal used in our measurements. As shown in Fig. S4(c), the molar ratio of the 4 elements at various points of the big crystal only shows small fluctuations in the ~8 mm × 8 mm single crystal, which means a large crystal cut from one position (the meaning of position is shown in Fig. S3) is quite homogeneous. Small crystals cut from this 8mm × 8mm big crystal show very similiar transport behaviors.

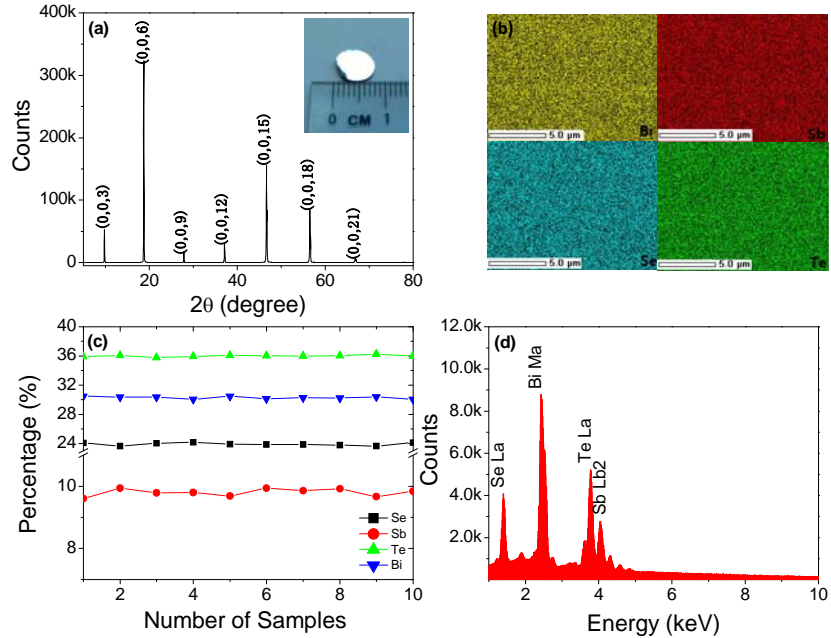

Fig. S4. Summary of structural characterization. (a) shows the wide angle x-ray diffraction from a bulk crystal oriented with the scattering vector perpendicular to the (100) family of planes. Inset is a photograph of a typical crystal. (b) The EDS mapping of a small area in a BSTS single crystal. (c) Molar ratio of Bi, Sb, Te, and Se at ten random positions in a big crystal like the one shown in the inset of (a). (d) A typical EDS spectroscopy.

The molar ratio of (Bi+Sb)/(Se+Te) and Bi/Sb in crystals cut from different positions (Fig. S3) of the crystal cylinder shows near-constant molar ratios of 2/3 and 3/1, respectively, while the Te/Se ratio changes from ~1.7/1.3 to ~1.9/1.1 for crystals obtained from different positions. Transport measurements show that, although the molar ratio of Te/Se varies with position as shown in Fig. S3, all the crystals show similar semiconducting $\rho(T)$ curve at high temperature and saturation behavior at low temperature and large resistivty (>1 $\Omega$cm) at 10 K. The value of $\rho(10\ K)/\rho(300K)$ can vary from 30 to 250. This characteristic of BSTS is very different from the condition of grown $Bi_2Te_3$ single crystals. $Bi_2Te_3$ single crystals obtained from different part of the cylinder usually show different transport behaviors, from metallic to semiconducting. We choose BSTS crystals with $\rho(10\ K)/\rho(300\ K) > 100$ and $\rho(10\ K) > 3\ \Omega\cdot cm$ for our experiments.

Combining the EDS results and the systematic decrease of the resistivity of nanoflake devices with decreasing thickness (the inset of Fig. 1(b) and Fig. S2), we can conclude that the resistivity change with thickness is not due to chemical inhomogeneity.

IV. Typical devices used for measurements.

All the devices were fabricated using mechanically exfoliated BSTS single crystal nanoflakes. The thickness of the nanoflakes varies from ~50 nm to ~600 nm. Standard photolithography technique was employed in the devices fabrication. Cr/Au contacts were deposited using a high vacuum sputtering system with a base pressure of $1 \times 10^{-8}$ Torr.

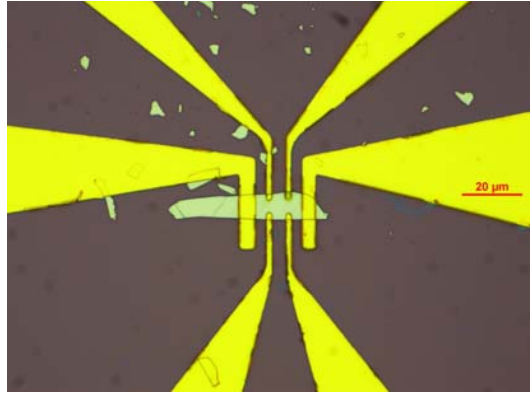

Fig. S5, Image of a typical electric gating devices for resistance and Hall resistance measurements

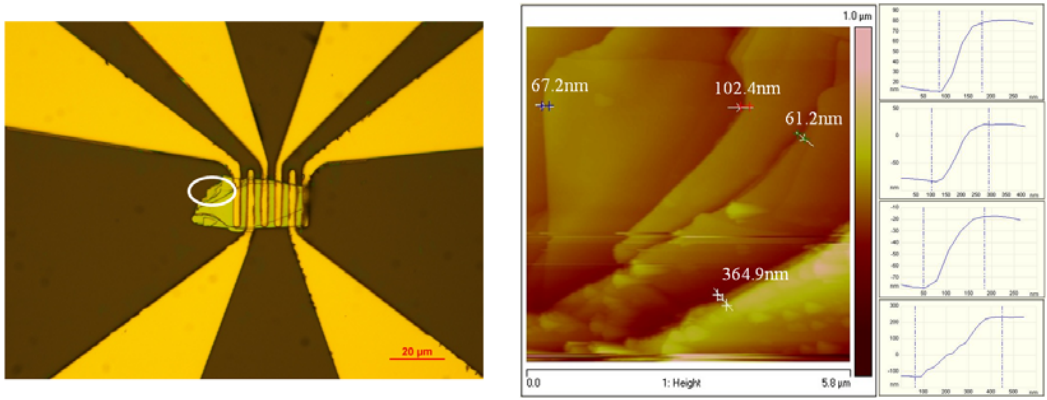

Fig. S6. Image of s typical device for weak antilocalization, angle dependent magnetoresistance, and universal fluctuation conductance measurements. Fig. 3 and Fig. 4 are the measurement results of this device. The thickness of the nanoflake used in this device is 596 nm. There are four steps from the bottom to the top in the circled region. The thickness of the sample is the total height of the four steps as shown in the AFM image.

## V. Time-dependent universal conductance fluctuation

We also observed a novel conductance fluctuation phenomenon in the nanoflake device under both low and high magnetic fields. The low magnetic field conductance fluctuation is very clear as shown in Fig. 3(b) and 3(f). Fig. S7(a) shows the conductance fluctuations at different temperatures below 10 K and between 4 and 9 Tesla perpendicular fields. Different from normal universal conductance fluctuations (UCF) discovered in topological insulator $Bi_2Se_3$ [22, 31], the conductance fluctuation in our nanoflake device evolves with time. Such time-dependent UCF can only be observed below 10 K and the magnitude of those fluctuations increases with decreasing temperature. The top 3 curves shown in Fig. S7(a) were measured continuously with different field sweeping direction. As shown in the circled region in the figure, the curves show very similar fluctuation behaviors in the same magnetic field region because the time interval between the measurements is short. With the time evolution, the fluctuation patterns become different in the same magnetic field region. Time-dependent UCF is due to the sensitivity of the conductance to the motion of individual scatters and such phenomena have been reported. The variation of conductance pattern is due to the motion of scattering sites which changes the interference pattern of all the intersecting electronic paths in the coherent volume of the scattering sites. Since the motion of scattering sites is related to the time, the observed fluctuations are also time-dependent which allows the non-retraceable results. As the electron transport in the nanoflake is almost fully surface transport as discussed before, the motion of the scattering site might be due to the time dependent surface contamination in the measurement chamber. Each magnetoresistancecurve from 4 to 9 Tesla takes about 1 hour. According to the results, we can conclude that the scattering site pattern on the topological surface changes after about 30 minutes.

The amplitude of fluctuation detected is around $0.2e^2/h$. The phase coherence length of electron $L_\varphi$ observed at 2 K is ~0.18 μm (Fig. 4) and the thermal length, $L_T = \sqrt{hD/k_BT}$ is around 1μm at 2 K. As $L_\varphi < L_T < L$, the fluctuation of $G$ can be calculated as $\delta G = e^2/h [L_\varphi / L]^{4-d}$, where L is the sample size. The calculated $\delta G$ is $0.3e^2/h$ which is quite close to our experiment data $0.2e^2/h$ at 2 K. Furthermore, the root-mean-square (rms) of $\delta G$ vs. temperature curve is also plotted as shown in Fig. S7(b). The rms of $\delta G$ is usually following the power law $rms\delta G \propto T^{-0.5}$ for two dimensional system. Our fitting shows $rms\delta G \propto T^{-0.43}$, which also indicates that the time-dependent UCF observed in the BSTS device sample is mainly from the surface states.

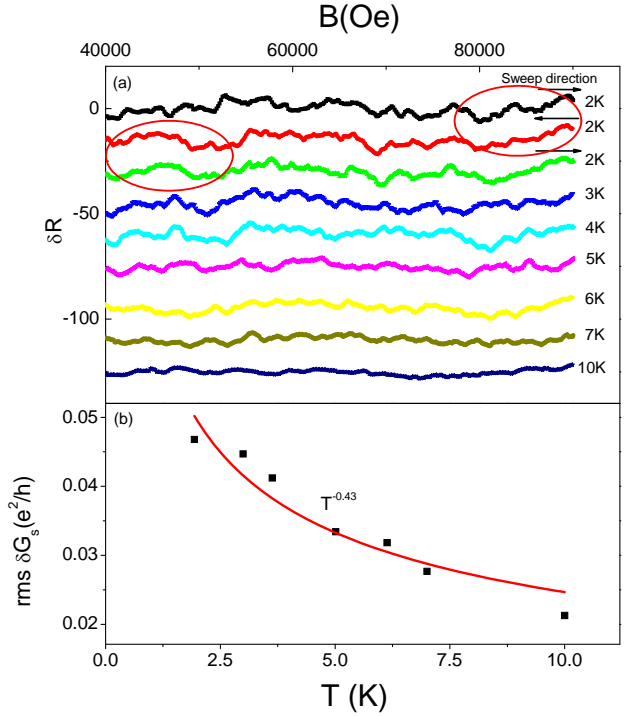

Fig. S7. (a) Magnetofingerprint (UCF) signal δR *vs* magnetic field for the nanoflake device. At 2 K, we scan the field for three times with alternative field sweeping direction (the arrows). Seven UCF curves measured at temperatures between 2 K to 10 K were shown in the figures. (b) Temperature dependence of root mean square of the fluctuation of conductance. The solid line is the fitting curve using power law and the fitted power is -0.43.

VI. More results for gated transport measurements and weak antilocalization measurements

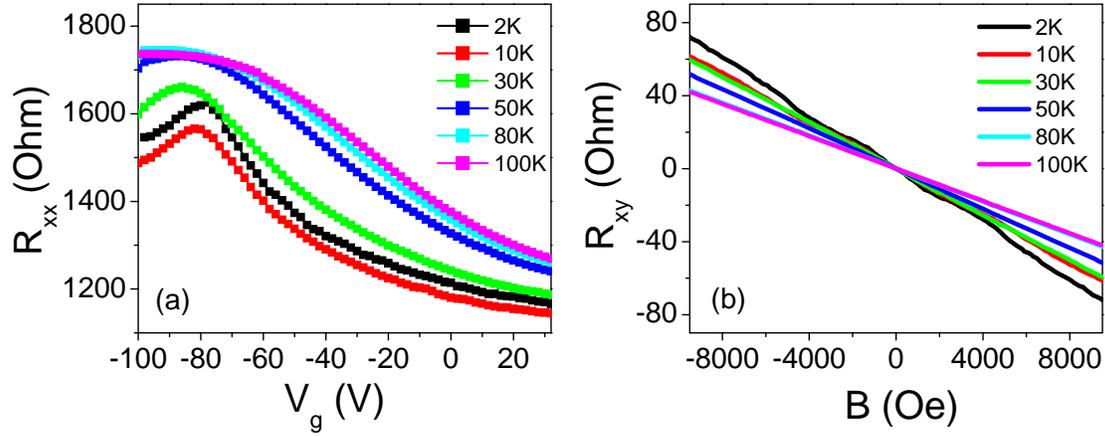

Fig. S8. (a) The $R_{xx}(V_g)$ curves of a BSTS nanoflake devices at various temperatures. The devices had been stored in vacuum for 2 months. The peak shifts to -80 volts. (b) The low field Hall resistance of the device.

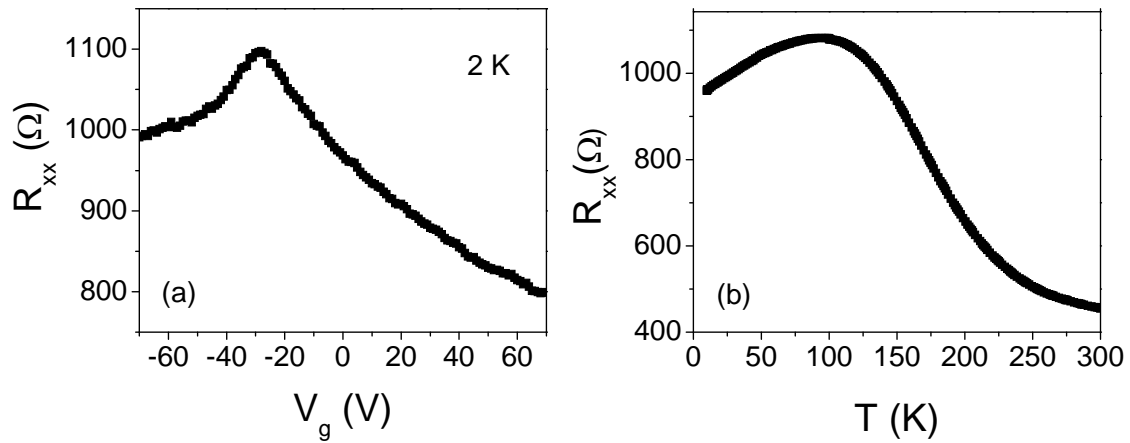

Fig S9. (a) The $R_{xx}(V_g)$ curves of a 212 nm thickness BSTS nanoflake devices at 2 K. (b) The $R_{xx}(T)$ curve of the device.

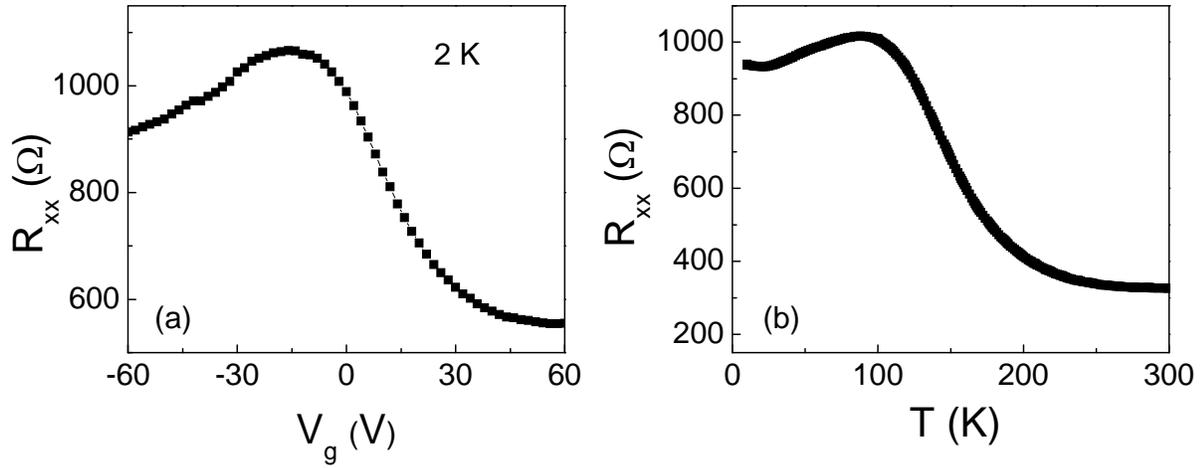

Fig. S10. (a) The $R_{xx}(V_g)$ curves of a 142 nm thickness BSTS nanoflake devices at 2 K. (b) The $R_{xx}(T)$ curve of the device.

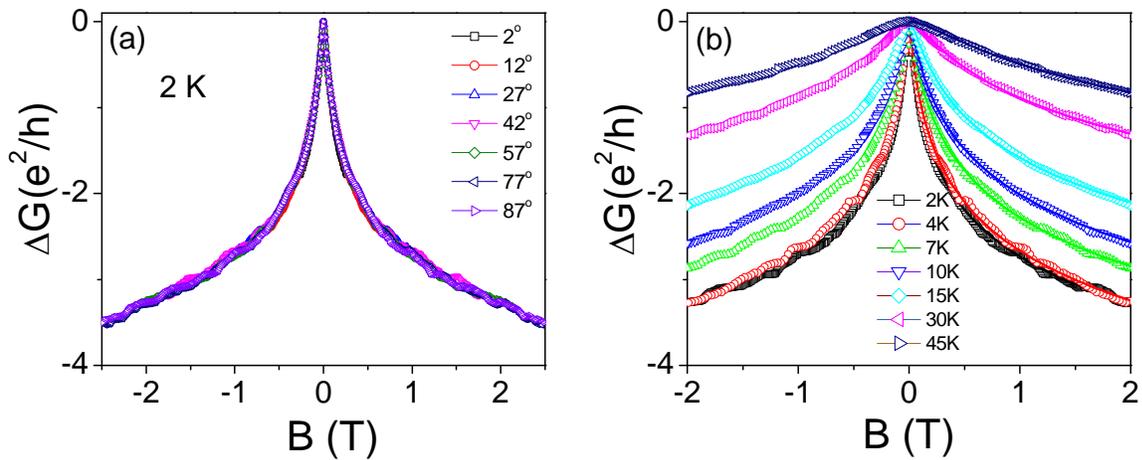

Fig. S11. (a) The weak antilocalization curves of a 198 nm thick devices fitted to a universal curve when the x axis is the perpendicular part of the magnetic field. (b) The weak antilocalization curve of the device at various temperatures.